# Quasi Regular Polyhedra and Their Duals with Coxeter Symmetries Represented by Quaternions I

Mehmet Koca[a)] and Nazife Ozdes Koca[b)]

Department of Physics, College of Science, Sultan Qaboos University
P.O. Box 36, Al-Khoud, 123 Muscat, Sultanate of Oman

Ramazan Koc[c)]
Department of Physics, Gaziantep University, 27310, Gaziantep, Turkey

**Abstract**

In two series of papers we construct quasi regular polyhedra and their duals which are similar to the Catalan solids. The group elements as well as the vertices of the polyhedra are represented in terms of quaternions. In the present paper we discuss the quasi regular polygons (isogonal and isotoxal polygons) using 2D Coxeter diagrams. In particular, we discuss the isogonal hexagons, octagons and decagons derived from 2D Coxeter diagrams and obtain aperiodic tilings of the plane with the isogonal polygons along with the regular polygons. We point out that one type of aperiodic tiling of the plane with regular and isogonal hexagons may represent a state of graphene where one carbon atom is bound to three neighboring carbons with two single bonds and one double bond. We also show how the plane can be tiled with two tiles; one of them is the isotoxal polygon, dual of the isogonal polygon. A general method is employed for the constructions of the quasi regular prisms and their duals in 3D dimensions with the use of 3D Coxeter diagrams.

[a)] electronic-mail: kocam@squ.edu.om
[b)] electronic-mail: nazife@squ.edu.om
[c)] electronic-mail: koc@gantep.edu.tr

## I. INTRODUCTION

The discovery of graphene [1], an infinite sheet of carbon atoms [2], tiled with regular hexagons invokes further investigations regarding the tiling of the plane with triangular symmetries suitable for carbon bondings including the double bonds with neighboring carbon atoms. Graphene has always been studied in the state that the carbon atoms form a honeycomb lattice where the edges of the regular hexagons represent the single bonds with the neighboring atoms and the fourth valence electron is uncoupled. Of course, there is no a convincing argument that the carbon may not form a double bond with one of the neighboring carbon atom to form a neutral state. When this happens, the double bond will be slightly shorter than the single bond leading to a deformation of the regular hexagon. Then a natural question arises as to whether it is possible to tile the Euclidean plane suitable for this type of bonding. The answer is yes. This was the main motivation for us to work on the constructions of the quasi regular polygons in 2D dimensions with Coxeter symmetries. The quasi regular polygons are of two types: the isogonal polygons consisting of two alternating unequal edges with equal interior angles; the isotoxal polygons consisting of equal edges but with alternating unequal interior angles. The isotogonal polygon with 2n sides is vertex transitive under the dihedral symmetry $D_n$. Its dual polygon, the isotoxal polygon, is edge transitive under the same symmetry. Aperiodic tiling of the plane with the isogonal and the isotoxal polygons is an interesting problem by itself. As we will see in this paper that the aperiodic tiling of the plane by two tiles, the isogonal hexagons and the regular hexagons is possible which could be considered a state of graphene if two bond lengths are chosen properly. Whether this phenomenon materializes in a graphene state is not important from the point of view of the study of the quasi regular polyhedra possessing the Coxeter symmetries. The topic deserves investigations because it seems that it has not been worked out in the literature in the context of Coxeter symmetries. In addition to the quasi regular polygons we also work out the structures of the quasi regular prisms and their dual solids possessing the Coxeter symmetries $D_n \times C_2$. The paper is organized as follows.

In section II we introduce the Coxeter diagrams for 2D and 3D dimensions [3] describing the root spaces as well as their dual spaces. The roots and the reflection generators with respect to the planes are represented by quaternions [4]. Earlier we have constructed the regular and Archimedean 3D polyhedra and their duals, the Catalan solids with quaternions [5]. In section III we construct some even-sided polygons with alternating two edge lengths (isogonal polygons) having equal interior angles. We also construct their dual polygons (isotoxal polygons) with equal edge lengths but with alternating two interior angles. These quasi regular polygons possess the Coxeter symmetries obtained from the 2D Coxeter diagrams. We work, in particular, with the three-fold, four-fold and the five-fold Coxeter symmetries and construct regular as well as quasi regular polygons invariant under these symmetries. We give tilings of the plane with isogonal and isotoxal hexagons, octagons and decagons with the constraint that at each vertex we have two quasi regular polygons together. With this approach we observe that in the case of $W(A_2) \approx D_3$ symmetry the Euclidean plane can be tiled by two isogonal hexagons having two different edge lengths and a regular hexagon all sharing the same vertex. This configuration may be regarded as the neutral graphene state where the long edges represent the single bonds and short edges represent the double bonds. The quasi regular polygons with dihedral symmetries $W(B_2) \approx D_4$ and $W(H_2) \approx D_5$ are also constructed and the tiling of the plane with the isogonal and isotoxal octagons and decagons are discussed. In section IV we extend algebraic approach to the constructions of the quasi regular prisms and their duals. The method is such that the regular prisms and their duals can be reproduced with a suitable choice of a relative scale factor. The group elements and the vertices are represented by quaternions. Section V is devoted to the concluding remarks.

## II. 2D AND 3D COXETER DIAGRAMS WITH QUATERNIONS

All Coxeter diagrams are represented with one type of simple roots [3] contrary to the Dynkin diagrams representing the Lie algebra root systems having long and short roots. In 2D space we have an infinite number of Coxeter diagrams shown in **Fig.1**. It is customary to use the notations $I_2(n)$ for the 2D Coxeter diagrams however we will continue using the notations $A_2$, $B_2$, $H_2$ for $I_2(n)$, when $n = 3, 4, 5$ respectively.

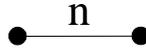

**FIG. 1. The Coxeter diagram $I_2(n)$ for 2D dimensions.**

Here $n$ represents the integers with $n = 2, 3, 4, 5, \ldots$. This means that the angle between the simple roots $\alpha_1$ and $\alpha_2$ is $\frac{(n-1)\pi}{n}$. We choose the norm of the roots $\sqrt{2}$ to be consistent with the Dynkin diagrams when they coincide. The Cartan matrix defined by the scalar product $C_{ij} = (\alpha_i, \alpha_j)$ of the 2D diagrams and its inverse $(C^{-1})_{ij} = (\omega_i, \omega_j)$, $(i, j = 1, 2)$ will read

$$C = \begin{bmatrix} 2 & 2\cos\frac{(n-1)\pi}{n} \\ 2\cos\frac{(n-1)\pi}{n} & 2 \end{bmatrix},$$

$$C^{-1} = \frac{1}{4\sin^2\frac{(n-1)\pi}{n}} \begin{bmatrix} 2 & -2\cos\frac{(n-1)\pi}{n} \\ -2\cos\frac{(n-1)\pi}{n} & 2 \end{bmatrix}. \quad (1)$$

The fundamental weights $\omega_i$ are the basis vectors of the dual space defined by the relation $(\alpha_i, \omega_j) = \delta_{ij}$ [6] where $\delta_{ij}$ is the Kronecker delta. The simple roots and the fundamental weights are related to each other by the relations:

$$\omega_i = (C^{-1})_{ij}\alpha_j, \quad \alpha_i = C_{ij}\omega_j. \tag{2}$$

Summation over the repeated index is implicit. Action of the reflection generator $r_i$ on an arbitrary vector $\Lambda$ is defined by the relation

$$r_i\Lambda = \Lambda - (\Lambda,\alpha_i)\alpha_i \quad (\text{no summation over i}) \tag{3}$$

They generate the dihedral group $D_n$ of order $2n$ satisfying the relations $r_1^2 = r_2^2 = (r_1 r_2)^n = 1$.

The groups generated by reflections which represent the symmetries of the regular and quasi regular prisms are given by the Coxeter diagrams depicted in **Fig. 2**.

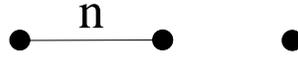

**FIG. 2. The Coxeter diagrams for prisms**

The symmetry group of the prisms and the quasi prisms are given by the group $D_n \times C_2$ of order $4n$. The group elements of the 2D, 3D and 4D Coxeter groups can be represented by the quaternions [4-5]. There is an interesting relation between the finite subgroups of quaternions and the Coxeter groups in 2D, 3D and 4D dimensions.

Let $q = q_0 + q_i e_i$, $(i = 1,2,3)$ be a real unit quaternion with its conjugate defined by $\bar{q} = q_0 - q_i e_i$ and the norm $q\bar{q} = \bar{q}q = 1$. Here the quaternionic imaginary units satisfy the relations

$$e_i e_j = -\delta_{ij} + \varepsilon_{ijk} e_k, \quad (i,j,k = 1,2,3) \tag{4}$$

where $\varepsilon_{ijk}$ is the usual Levi-Civita symbol and summation over the repeated indices is implicit. The correspondence between the usual quaternionic imaginary units and ours are given by $i = e_1$, $j = e_2$, $k = e_3$. We define the scalar product of two real quaternions as

$$(p,q) = \frac{1}{2}(\bar{p}q + \bar{q}p). \tag{5}$$

If we choose the simple roots in terms of the quaternions $\alpha_1 = \sqrt{2}$, $\alpha_2 = \sqrt{2}\exp[e_1(n-1)\pi/n]$ and $\alpha_3 = \sqrt{2}e_2$ one can obtain the Cartan matrices for the dihedral groups given in (1) as the sub-matrices of the Cartan matrix for the prismatic symmetries which is given by

$$C = \begin{bmatrix} 2 & 2\cos\dfrac{(n-1)\pi}{n} & 0 \\ 2\cos\dfrac{(n-1)\pi}{n} & 2 & 0 \\ 0 & 0 & 2 \end{bmatrix}. \tag{6a}$$

Its inverse is given as

$$C^{-1} = \begin{bmatrix} \dfrac{1}{2\sin^2\dfrac{(n-1)\pi}{n}} & \dfrac{-\cos\dfrac{(n-1)\pi}{n}}{2\sin^2\dfrac{(n-1)\pi}{n}} & 0 \\ \dfrac{-\cos\dfrac{(n-1)\pi}{n}}{2\sin^2\dfrac{(n-1)\pi}{n}} & \dfrac{1}{2\sin^2\dfrac{(n-1)\pi}{n}} & 0 \\ 0 & 0 & \dfrac{1}{2} \end{bmatrix}. \tag{6b}$$

When $\Lambda$ and $\alpha_i$ are represented by quaternions the equation (3) can be written as

$$r_i \Lambda = -\frac{1}{2}\alpha_i \bar{\Lambda} \alpha_i \equiv \frac{1}{2}[\,\alpha_i\,,-\alpha_i]^*\Lambda \tag{7}$$

or abstractly we define the generators [4],

$$r_i \equiv [\,\frac{\alpha_i}{\sqrt{2}}\,,-\frac{\alpha_i}{\sqrt{2}}]^*\,. \tag{8}$$

Here the roots are complex numbers for 2D diagrams, a subset of quaternions, where the unit complex numbers $\alpha_1/\sqrt{2}$ and $\alpha_2/\sqrt{2}$ generate the cyclic group of order $n$; however, the generators in (8) generate the dihedral group of order $2n$. Similarly the unit quaternions $\alpha_1/\sqrt{2}$, $\alpha_2/\sqrt{2}$ and $\alpha_3/\sqrt{2}$ generate the dicyclic group of order $2n$ but the reflection generators $r_1$, $r_2$ and $r_3$ generate the group $D_n \times C_2$ of order $4n$. The root system of the 2D Coxeter diagram can be constructed in a simple manner in terms of quaternions. Let $q = \exp(e_1 \dfrac{\pi}{n})$ be the unit quaternion. Then all integer powers of $q, \{q^k, k = 1, 2, ..2n\}$, constitute a scaled copy of the root system of 2D Coxeter diagram. If $\Lambda$ represents any quaternion, the generators act on $\Lambda$ as follows:

$$r_1 \equiv [1, -1]^*, \quad r_1\Lambda = -\bar{\Lambda}; \quad r_2 \equiv [e_1^{\frac{(n-1)\pi}{n}}, -e_1^{\frac{(n-1)\pi}{n}}]^*, \quad r_2\Lambda = -e_1^{\frac{(n-1)\pi}{n}} \bar{\Lambda} e_1^{\frac{(n-1)\pi}{n}};$$

$$r_3 \equiv [e_2, -e_2], \quad r_3\Lambda = -e_2\bar{\Lambda}e_2. \tag{9}$$

## III. 2D COXETER DIAGRAMS AND QUASI REGULAR POLYGONS

To construct the weights of the irreducible representation of the Lie algebra one starts with the highest weight. If $g$ is the Lie algebra of rank $l$ then the highest weight

$$\Lambda = a_1\omega_1 + a_2\omega_2 + ... + a_l\omega_l \equiv (a_1, a_2, ..., a_l) \tag{10}$$

is represented by the $l$ non-negative integers [7]. Applying the Coxeter- Weyl group $W(g)$ on the highest weight one can generate the $\Lambda - orbit \equiv O(\Lambda) = W(g)\Lambda$. In order to obtain all the weights of a particular representation of the Lie algebra one should use the Weyl's [7] formula which involves several orbits. But the orbit $O(\Lambda)$ represents a single orbit describing a polytope (quasi regular in general) in $l$ dimensional Euclidean space. In 2D space the orbits of the Coxeter group is either regular polygons or even-sided quasi regular polygons. In what follows we will discuss the orbits of the Coxeter groups with $n = 2, 3, 4, 5$. We remark here that there does not correspond any Lie algebra for the Coxeter diagrams $n = 5$ and $n \geq 7$. Since we shall be dealing with the 2D diagrams, in what follows, it is appropriate to determine the vertices of the polygons in terms of complex numbers. The fundamental weights can be written as

$$\omega_1 = \frac{1}{\sqrt{2}}(1 - e_1 \frac{\cos\frac{(n-1)\pi}{n}}{\sin\frac{(n-1)\pi}{n}}), \quad \omega_2 = \frac{e_1}{\sqrt{2}\sin\frac{(n-1)\pi}{n}}. \tag{11}$$

A general vertex $\Lambda = a_1\omega_1 + a_2\omega_2$ is then an arbitrary complex number which can be written as

$$\Lambda = \frac{1}{\sqrt{2}}[a_1 + e_1 \frac{1}{\sin\frac{(n-1)\pi}{n}}(-a_1\cos\frac{(n-1)\pi}{n} + a_2)] \tag{12a}$$

and the generators act as follows,

$$r_1\Lambda = -\bar{\Lambda}, \quad r_2\Lambda = -e_1^{\frac{-2\pi}{n}}\bar{\Lambda}, \quad r_1r_2\Lambda = e^{\frac{2\pi}{n}e_1}\Lambda, \quad r_2r_1\Lambda = e^{-\frac{2\pi}{n}e_1}\Lambda. \tag{12b}$$

It is clear from (12b) that $r_1^2 = r_2^2 = (r_1r_2)^n = (r_2r_1)^n = 1$.

The *2n* vertices of the polygon can be determined as

$$(r_1 r_2)^k \Lambda = e^{\frac{2\pi k}{n} e_1} \Lambda, \quad (r_1 r_2)^k (r_1 \Lambda) = -e^{\frac{2\pi k}{n} e_1} \overline{\Lambda}, \quad k = 1, 2, ..., n. \tag{13}$$

For $a_1 = a_2 = a$ the 2n vertices of a regular polygon of edge length $\sqrt{2}a$ are given in terms of complex numbers by a simple formula

$$\frac{a}{\sqrt{2}\cos(\frac{(n-1)\pi}{n})} e^{e_1 \frac{\pi}{2n}(n-1+4k)}, \quad \frac{-a}{\sqrt{2}\cos(\frac{(n-1)\pi}{n})} e^{e_1 \frac{\pi}{2n}(1-n+4k)}, \quad k = 1, 2, ...n. \tag{14}$$

Now we deal with the special cases.

**A. *n*=2 with $D_2 = C_2 \times C_2$ symmetry**

Two simple roots $\alpha_1 = \sqrt{2}$ and $\alpha_2 = \sqrt{2}e_1$, representing the Coxeter diagram $A_1 \oplus A_1$, are orthogonal to each other where the generators form the Klein's four-group $D_2 = C_2 \times C_2$. A general orbit is obtained by acting the group elements on the vector $\Lambda = \frac{1}{\sqrt{2}}(a_1 + e_1 a_2)$.

Since we are not dealing with the Lie algebra it is not necessary to restrict the values of $a_i$ to non-negative integers. They can be any real number in general. However, we will follow the highest weight technique when the Coxeter diagrams coincide with the Dynkin diagrams. For $\Lambda = \omega_1 \equiv (10)$ and $\Lambda = \omega_2 \equiv (01)$, the orbits are two segments of straight lines perpendicular to each other. The orbit $O(\Lambda = \omega_1 + \omega_2) \equiv O(11)$ involves the vectors $\pm \omega_1 = \pm \frac{1}{\sqrt{2}}, \pm \omega_2 = \pm \frac{1}{\sqrt{2}} e_1$ which form a square. For all 2D Coxeter diagrams when the orbit is derived from the vector $\Lambda = a(\omega_1 + \omega_2) \equiv a(11)$, where $a$ is an arbitrary real number, then the polygon has an additional symmetry. It is the symmetry of the Coxeter-Dynkin diagram which can be defined by the generator $\gamma : \alpha_1 \leftrightarrow \alpha_2$ leading to a larger symmetry $W(g) : C_2$ where (: ) denotes the semi-direct product of two groups. In the above case the group is $D_4 \approx (C_2 \times C_2) : C_2 \approx C_4 : C_2'$ of order 8.

When we consider the most general case, namely, $\Lambda = a_1 \omega_1 + a_2 \omega_2$ the orbit represents a rectangle of sides $\sqrt{2}a_1$, $\sqrt{2}a_2$. The rectangle is an isogonal polygon with the $D_2$ symmetry. The dual of the rectangle is a rhombus (an isotoxal polygon) whose vertices can be determined by taking the mid point of one of the edge of the rectangle, say, the vector $a_1 \omega_1$. We take the other vector $\lambda \omega_2$ bisecting the edge of length $\sqrt{2}a_1$ of the rectangle. To determine the dual of the rectangle the line joining these vectors $\lambda \omega_2 - a_1 \omega_1$ must be orthogonal to the vector $\Lambda = a_1 \omega_1 + a_2 \omega_2$ which determines the scale factor $\lambda = \dfrac{a_1^2}{a_2}$. The vertices which consist of two fundamental orbits $O(a_1 0) = \{\pm a_1 \omega_1\}$ and $\lambda O(10) = \{\pm \lambda \omega_2\}$ represent a rhombus. The ratio of lengths of the diagonals of the rhombus is $\dfrac{a_1}{a_2}$. The rectangle is vertex transitive under the Klein's group $C_2 \times C_2$ and its dual rhombus is edge transitive. For the values $a_1 = 1$ and $a_2 = 2$ the rectangle and its dual rhombus are given in **Fig. 3**.

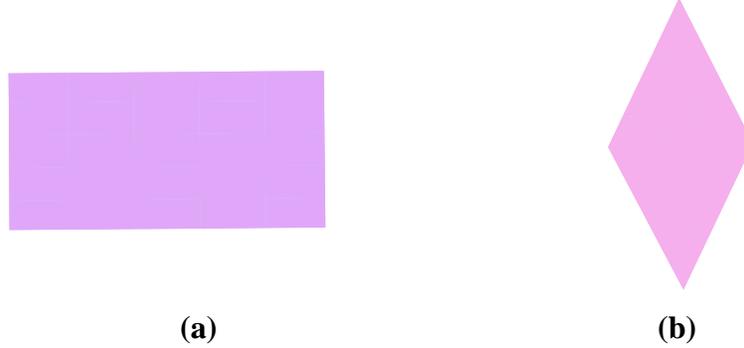

(a)          (b)

**FIG. 3. The rectangle (a) and its dual rhombus (b) possessing the symmetry $D_2 = C_2 \times C_2$.**

**B.** $n=3$ with $W(A_2) \approx D_3 \approx S_3$ **symmetry**

The Coxeter group $W(A_2) \approx D_3 \approx S_3$ consists of six elements. The three elements $r_1$, $r_2$, $r_1 r_2 r_1 = r_2 r_1 r_2$ represent reflections with respect to the lines orthogonal to the roots $\alpha_1$, $\alpha_2$, $\alpha_1 + \alpha_2$ respectively and the rotational group elements $1$, $r_1 r_2$ and $(r_1 r_2)^2$ represent the cyclic group $C_3$ which rotates the system by $120^0$. We have two fundamental orbits $O(10) = \{\omega_1, \omega_2 - \omega_1, -\omega_2\}$ and $O(01) = \{\omega_2, -\omega_1, \omega_1 - \omega_2\}$. Each orbit represents an equilateral triangle, dual to each other which are transformed to each other by the Coxeter-Dynkin diagram symmetry.

The orbit $O(11) = O(\omega_1 + \omega_2)$ represents a regular hexagon which has a larger symmetry because of the diagram symmetry $\gamma : \alpha_1 \leftrightarrow \alpha_2$. Then we obtain the full symmetry group $D_3 : C_2 \approx D_6$ of the regular hexagon. The regular hexagonal orbit can also be obtained from the Coxeter diagram with $n = 6$ by applying the $D_6 \approx W(G_2)$ generators on the either weight $O(10) = W(G_2)\omega_1$ or $O(01) = W(G_2)\omega_2$ where $\omega_1$ and $\omega_2$ are the weights of the Coxeter diagram $G_2$. Now we discuss the orbit obtained from the vector $\Lambda = a_1 \omega_1 + a_2 \omega_2$ where $a_1 \neq a_2 \neq 0$. The Coxeter group $W(A_2)$ generates the six elements of the orbit $O(\Lambda) = O(a_1 a_2)$ arranged in the counter clockwise order as follows:

$$O(a_1, a_2) = \{\Lambda_1, \Lambda_2, \Lambda_3, \Lambda_4, \Lambda_5, \Lambda_6\},$$

where

$$\begin{aligned}\Lambda_1 &= a_1\omega_1 + a_2\omega_2, \quad \Lambda_2 = -a_1\omega_1 + (a_1 + a_2)\omega_2, \quad \Lambda_3 = -(a_1 + a_2)\omega_1 + a_1\omega_2 \\ \Lambda_4 &= -a_2\omega_1 - a_1\omega_2, \quad \Lambda_5 = a_2\omega_1 - (a_1 + a_2)\omega_2, \quad \Lambda_6 = (a_1 + a_2)\omega_1 - a_2\omega_2\end{aligned} \quad (15)$$

They can also be obtained as complex numbers from (13) by substituting n=3 in (11). These vectors represent the vertices of an isogonal hexagon with $120^0$ interior angles and the alternating edges of lengths $\sqrt{2}a_1$ and $\sqrt{2}a_2$.

The dual of the isogonal hexagon is an isotoxal hexagon in which the edges are equal, however, it has two different alternating interior angels $\alpha$ and $\beta$ such that $\alpha + \beta = 240^0$. The vertices of the isotoxal hexagon lie on two fundamental orbits $O(a_1 0)$ and $\lambda O(01)$. The scale factor $\lambda$ is determined by the relation

$$(\lambda\omega_2 - a_1\omega_1).(a_1\omega_1 + a_2\omega_2) = 0 \quad \Rightarrow \quad \lambda = \frac{a_1(2a_1 + a_2)}{a_1 + 2a_2}. \quad (16)$$

Let $a_1\omega_1$ represents the center of one of the edge of the isogonal hexagon. The vertices of the isotoxal hexagon can then be written in the counter clockwise order as follows:

$$\{B_1 = a_1\omega_1, \; B_2 = \lambda\omega_2, \; B_3 = a_1(-\omega_1 + \omega_2), B_4 = -\lambda\omega_1, \; B_5 = -a_1\omega_2, \; B_6 = \lambda(\omega_1 - \omega_2)\}. \quad (17)$$

Defining $\eta = \dfrac{\lambda}{a_1}$ one can check that the edge length of the isotoxal hexagon is given by

$$a_1[\frac{2}{3}(\eta^2 - \eta + 1)]^{\frac{1}{2}}. \quad (18)$$

We have two different interior angels given by

$$\alpha = \cos^{-1}\frac{2\eta^2 - 2\eta - 1}{2(\eta^2 - \eta + 1)}, \quad \beta = \cos^{-1}\frac{-\eta^2 - 2\eta + 2}{2(\eta^2 - \eta + 1)}. \quad (19)$$

One can prove that for any value of $\eta$, $\alpha + \beta = 240^0$. The isogonal hexagon and its dual isotoxal hexagon are shown in **Fig. 4** for the values $a_1 = 1$ and $a_2 = 2$.

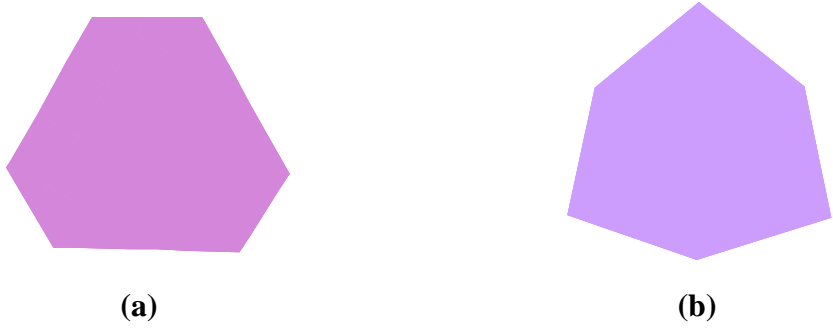

**FIG. 4. The isogonal hexagon (a) and its dual isotoxal hexagon (b) possessing the dihedral symmetry $D_3$.**

One can tile the plane with regular hexagons and isogonal hexagons with the constraint that two isogonal hexagons and one regular hexagon share the same vertex regardless of the values of $a_1$ and $a_2$. One type of tiling of the plane is shown in **Fig. 5(a)** with the values $a_1 = 1$ and $a_2 = 2$. Another tiling is shown in **Fig. 5(b)** for the isogonal hexagon for $a_1 = 2$ and $a_2 = 1$. In the limit $a_1 \to 0$ and $a_2 = 1$ isogonal hexagon turns out to be an equilateral triangle and such tiling is depicted in **Fig. 5(c)**. The other extreme limit is shown in **Fig. 5(d)** where $a_1 = 1$ and $a_2 \to 0$. The honeycomb lattice corresponds to the case where $a_1 = a_2$ which is shown in the **Fig. 5(e)**. The honeycomb lattice representing the tiling of the plane with regular hexagons possesses translational invariance, that is to say, it is invariant under the affine Coxeter group $\hat{W}(A_2)$. However the tilings represented by **Fig.5 (a-d)** violate the translational invariance. Such tilings are said to be aperiodic tiling representing a quasi crystal lattice [8]. We anticipate that a neutral state of the graphene consisting of infinite number of carbon atoms can be represented by a tiling similar to the one in **Fig.5(a)** provided one of the parameters represents the double bonds (say $a_1$) and the other is representing the single bonds ($a_2$). Experimentally one expects $a_1 \prec a_2$ but nevertheless they are nearly equal each other contrary to the isogonal hexagon in **Fig.5(a)** which is an exaggerated version of this quasi crystal lattice. As we will discuss in the second paper [9] the $C_{60}$ molecule represents tiling of the sphere with isogonal hexagons and the pentagons for it has two bond lengths.

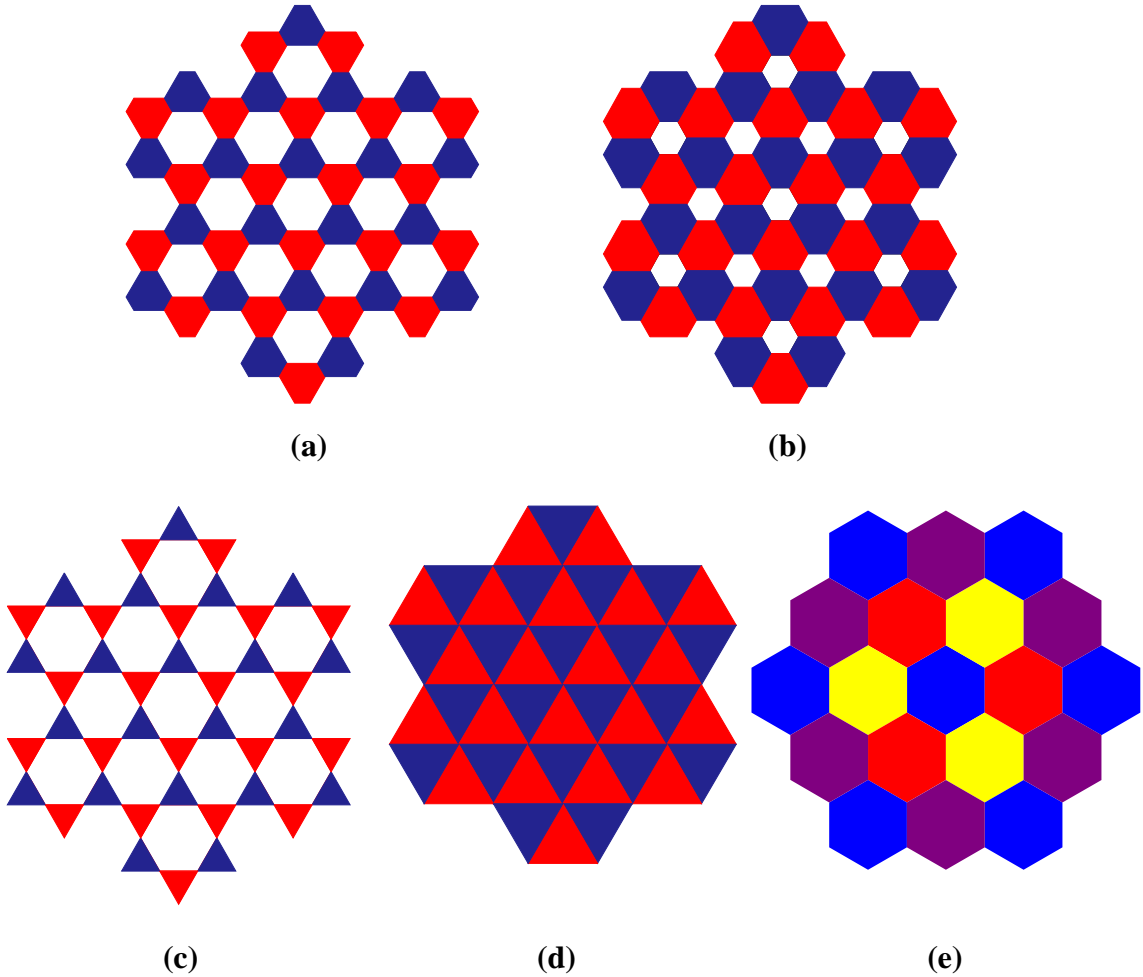

**(a)** **(b)**

**(c)** **(d)** **(e)**

**FIG. 5. Tiling of the plane with isogonal hexagons-regular hexagons (a-b), regular hexagons and triangles (c), tiling with triangles (d) and the honeycomb lattice (e)**

Tiling of the plane now can also be made with the isotoxal hexagons by joining its two vertices with two alternating angles $\alpha$ and $\beta$ so that one obtains an exterior angle of $120^0$ at each vertex. This way we create a regular hexagon surrounded by six isotoxal hexagons as shown in **Fig. 6**. This aperiodic tiling is the dual of the tiling in **Fig. 5(a).** In this tiling all hexagons have the same edge lengths. However at each vertex the three angles satisfy, as expected, the relation $\alpha + \beta + 120^0 = 360^0$. Here again when $\eta \to 1$ we obtain the honeycomb lattice of **Fig. 5(e)**.

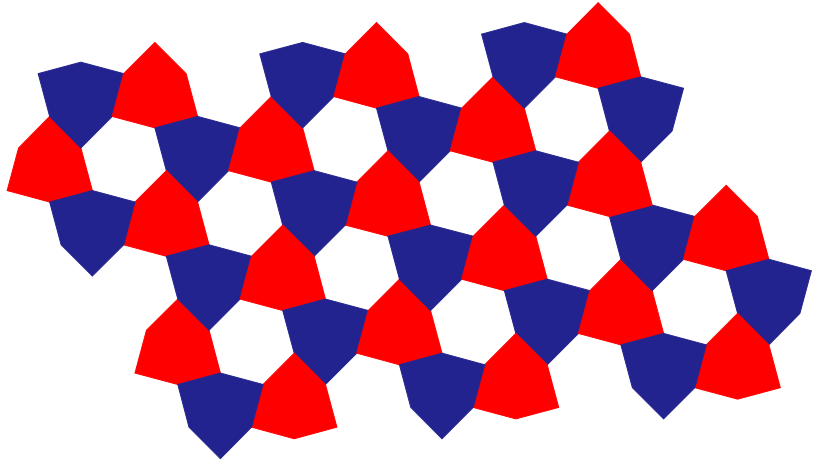

**FIG. 6. Tiling of the plane with isotoxal and regular hexagons.**

### C. $n = 4$ with $W(B_2) \approx D_4$ symmetry

We can repeat the similar arguments raised for the case n=3. Here also we have two fundamental orbits

$$O(10) = \{\omega_1, -\omega_1 + \sqrt{2}\omega_2, -\omega_1, \omega_1 - \sqrt{2}\omega_2\} \tag{20a}$$

$$O(01) = \{\omega_2, -\omega_2 + \sqrt{2}\omega_1, -\omega_2, \omega_2 - \sqrt{2}\omega_1\} . \tag{20b}$$

Each orbit represents a square.

The orbit $O(\omega_1 + \omega_2) = O(11)$ is a regular octagon possessing the symmetry $D_4 : C_2 \approx D_8$. One can tile the plane with regular octagon and the square as shown in **Fig.7**.

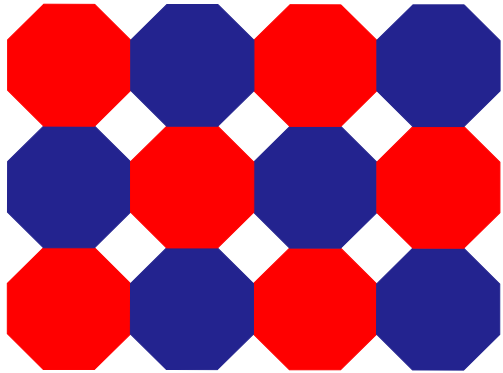

**FIG. 7. Tiling of the plane by regular octagons and squares**

The general orbit $O(\Lambda) = O(a_1 a_2)$ with $a_1 \neq a_2 \neq 0$ is an isogonal octagon with interior angles $135^0$. Isogonal octagons are vertex transitive under the dihedral group $D_4$. The vertices of the isogonal octagon can be computed in a simple manner as follows. The vertex $\Lambda = a_1 \omega_1 + a_2 \omega_2 = \frac{1}{\sqrt{2}}[a_1 + (a_1 + \sqrt{2} a_2) e_1]$ is a complex number and the reflection generators act on $\Lambda$ as $r_1 \Lambda = -\overline{\Lambda}$ and $r_2 \Lambda = e_1 \overline{\Lambda}$. We determine the vertices of the dual of the isogonal octagon by finding the scale factor $\lambda$ from the equation $(\lambda \omega_2 - a_1 \omega_1).(a_1 \omega_1 + a_2 \omega_2) = 0$ which leads to the value $\lambda = \frac{a_1(\sqrt{2} a_1 + a_2)}{(a_1 + \sqrt{2} a_2)}$.

Defining $\eta = \frac{\lambda}{a_1}$ we obtain the edge length of the isotoxal octagon given by the formula (18). The vertices of the isotoxal octagon is the union of the orbits $O(a_1 0)$ and $\lambda O(01)$. The alternating angles satisfy the relation $\alpha + \beta = 270^0$.

The isogonal octagon with values $a_1 = 1$ and $a_2 = 2$ and its dual isotoxal octagon are depicted in **Fig. 8.**

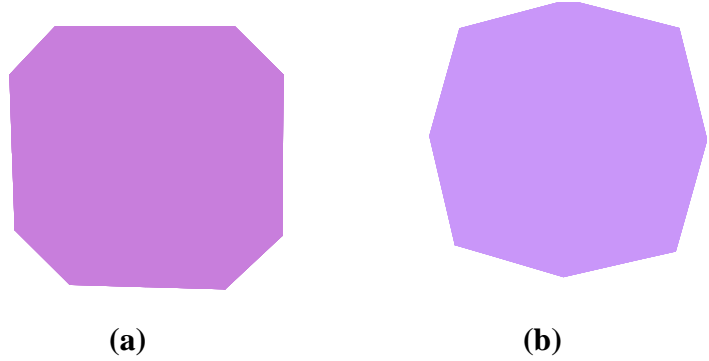

**(a)**                     **(b)**

**FIG. 8. The isogonal octagon (a) and its dual isotoxal octagon (b)**

The tiling of the plane with two isogonal octagons and one square at each vertex is shown in **Fig. 9(a)**. The tiling of the plane with two isotoxal octagons and one square at one vertex is shown in **Fig. 9(b)**.

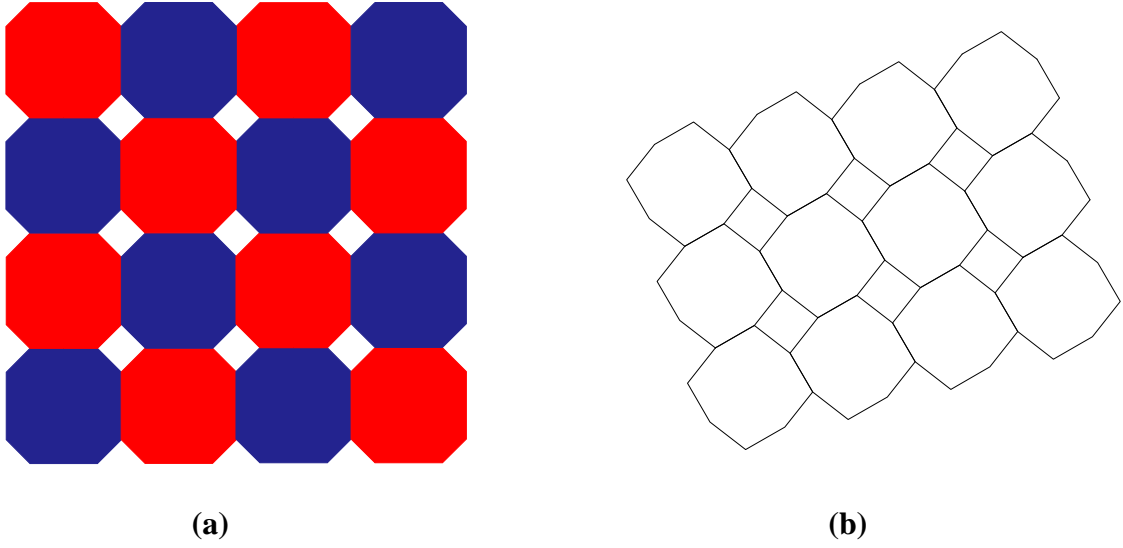

**(a)**                  **(b)**

**FIG. 9.  Aperiodic tiling of the plane by isogonal octagons (a) and isotoxal octagons (b) with squares**

### D. $n = 5$ with $W(H_2) \approx D_5$ symmetry

This case corresponds to the Coxeter group $W(H_2) \approx D_5$, the dihedral group of order 10. Here the Cartan matrix and its inverse corresponding to the Coxeter diagram $H_2$ can be written in terms of the golden ratio $\tau = \dfrac{1+\sqrt{5}}{2}$ and $\sigma = \dfrac{1-\sqrt{5}}{2}$ as

$$C = \begin{bmatrix} 2 & -\tau \\ -\tau & 2 \end{bmatrix} \text{ and } C^{-1} = \frac{1}{2+\sigma}\begin{bmatrix} 2 & \tau \\ \tau & 2 \end{bmatrix}. \tag{21}$$

The orbit of a general vector $\Lambda = a_1\omega_1 + a_2\omega_2$ can be obtained as the set of 10 vectors:

$$O(a_1,a_2) = \{\Lambda_1 = a_1\omega_1 + a_2\omega_2, \Lambda_2 = -a_1\omega_1 + (a_1\tau + a_2)\omega_2, \Lambda_3 = -(a_1+\tau a_2)\omega_1 + \tau(a_1+a_2)\omega_2,$$
$$\Lambda_4 = -\tau(a_1+a_2)\omega_1 + (a_1+\tau a_2)\omega_2, \Lambda_5 = -(\tau a_1+a_2)\omega_1 + a_1\omega_2, \Lambda_6 = -a_2\omega_1 - a_1\omega_2,$$
$$\Lambda_7 = a_2\omega_1 - (a_1+a_2\tau)\omega_2, \Lambda_8 = (\tau a_1+a_2)\omega_1 - \tau(a_1+a_2)\omega_2, \Lambda_9 = \tau(a_1+a_2)\omega_1 - (\tau a_1+a_2)\omega_2,$$
$$\Lambda_{10} = (a_1+\tau a_2)\omega_1 - a_2\omega_2\}.$$
$$\tag{22}$$

The fundamental orbits $O(10)$ and $O(01)$ are obtained by letting $a_1 = 1$, $a_2 = 0$ and $a_1 = 0$, $a_2 = 1$ respectively in (22). The fundamental orbits are the regular pentagons which are the duals of each other. The regular decagon is obtained by letting $a_1 = 1$ and $a_2 = 1$ in (22). The regular decagon is both vertex and edge transitive under the dihedral group $W(H_2):C_2 \approx D_{10}$ because of the diagram symmetry.

The aperiodic tiling of the plane with five-fold symmetry is introduced by Penrose [10] for the plane cannot be tiled only with regular pentagons. It has been recently shown that the Islamic tiling of the plane [11] with five-fold symmetry dates back to the medieval time. The Islamic architecture used five different tiles, decagon, pentagon, rhombus, nonregular hexagon and bow tie. Here we give in **Fig.10** one of those aperiodic tilings of the plane with regular decagons and bow ties. The tiling in **Fig.10** displays locally a dihedral symmetry $W(H_2) \approx D_5$ but it has no translational invariance.

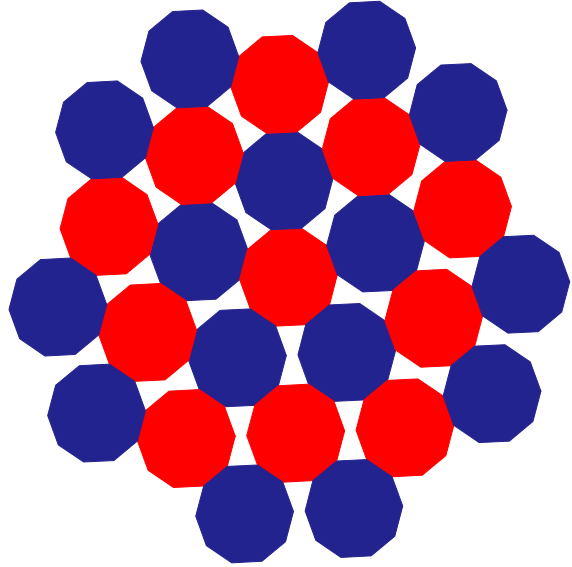

**FIG. 10.  The tiling of the plane by regular decagons and bow ties.**

One can also construct an isogonal decagon represented by alternating edge lengths $\sqrt{2}a_1$ and $\sqrt{2}a_2$ with $a_1 \neq a_2 \neq 0$ in (22). Its interior angles are all equal $144^0$. Note that the isogonal decagon is vertex transitive under the Coxeter symmetry $W(H_2) \approx D_5$. One such isogonal decagon is shown in **Fig. 11(a)** corresponding to the alternating edge lengths for $a_1 = 1$ and $a_2 = 2$. The dual of the isogonal decagon is the isotoxal decagon with equal edge lengths but with alternating angles $\alpha + \beta = 288^0$. Its vertices lie on two fundamental orbits of the Coxeter group $W(H_2) \approx D_5$, say, $O(a_1 0)$ and $O(0\lambda)$ where $\lambda$ is determined as usual given by $\lambda = \dfrac{a_1(2a_1 + \tau a_2)}{(\tau a_1 + 2a_2)}$. The edge length of the isotoxal decagon is given by the formula (18) where $\eta = \dfrac{(2a_1 + \tau a_2)}{(\tau a_1 + 2a_2)}$. The isotoxal decagon is shown in **Fig. 11(b).**

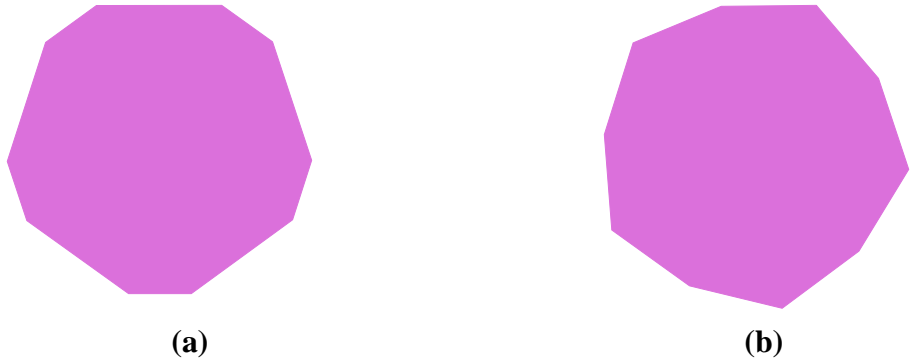

**FIG. 11.  The isogonal decagon (a) and its dual the isotoxal decagon (b)**

One type of aperiodic tiling of the plane with isogonal decagons is shown in **Fig.12.**

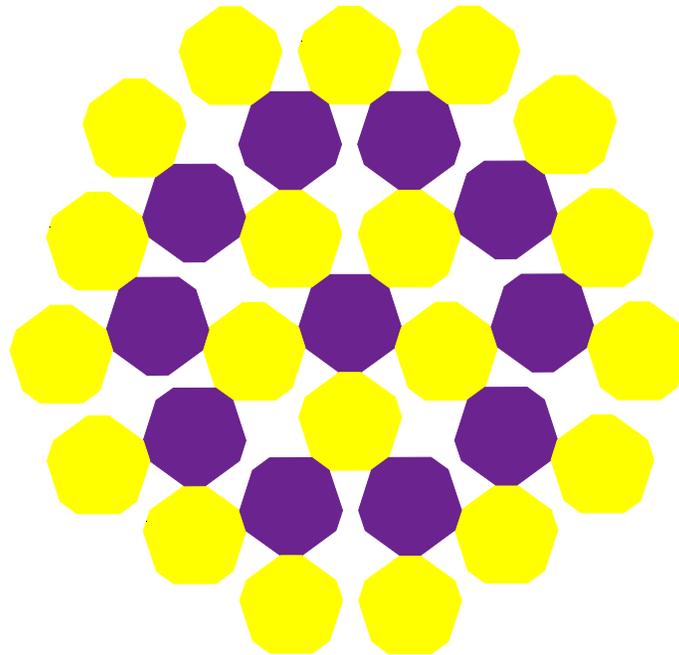

**FIG. 12. Tiling of the plane with isogonal decagons**

The above type of tilings of the plane can be extended for the isogonal polygons derived from the Coxeter symmetries for $n \geq 7$.

## IV. 3D COXETER DIAGRAMS AND QUASI REGULAR PRISMS

The Coxeter diagrams for the prismatic symmetries are discussed in Section II. In this section we discuss the constructions of the regular and quasi regular prismatic solids and their dual solids. Before we proceed further we note the following interesting aspect of the dicyclic groups of quaternions. Let $q = \exp(e_1 \frac{\pi}{n})$ be the unit quaternion of order $2n$. We have noted in Section II that the set of $2n$ quaternions obtained by taking all integer powers of $q$ constitute a scaled root system of the 2D Coxeter diagram of rank 2. Let us define the unit quaternion by $q' = \exp(e_1 \frac{\pi}{n})e_2$. It is clear from this definition that the new set of $2n$ quaternions constitute a scaled copy of another 2D Coxeter diagram of rank 2 orthogonal to the first one as shown in **Fig.13**.

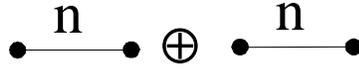

**FIG. 13.** The Coxeter diagram $I_2(n) \oplus I_2(n)$ **of rank-4.**

It is interesting to note that the elements of dicyclic group of the quaternions constitute the root system of this rank-4 Coxeter diagram. The automorphism group of the diagram in **Fig.13** can be represented as the set of elements constructed as pairs of quaternions, multiplying an arbitrary quaternion from left and right, selected randomly from the set of elements of the dicyclic group of quaternions. The root system of the rank-4 Coxeter diagram can be represented by the quaternions

$$\text{Root System of } I_2(n) \oplus I_2(n) = \{q^k, q'^k; \ k=1,2,...,2n\}. \tag{23}$$

Let $s$ and $t$ be an arbitrary element of the set of (23). Then the automorphism group of the set $I_2(n) \oplus I_2(n) \equiv \{q^k, q'^k; \ k=1,2,...,2n\}$ can be written as

$$Aut(I_2(n) \oplus I_2(n)) = \{[s,t] \oplus [s,t]^*\}, \quad s,t \in \{I_2(n) \oplus I_2(n)\}. \tag{24}$$

This is a group of order $4n \times 4n$. When $n = 3$ the automorphism group of (24) is of order 144 and for $n = 5$ it is of order 400. We have noted in an earlier paper [12] that both groups are the maximal subgroups of the Coxeter group $W(H_4)$ of order 14,400. It was also shown that the group $Aut(H_2 \oplus H_2) = \{[s,t] \oplus [s,t]^*\}$ is the symmetry group of the grand antiprism [13].

Here we are interested in rank-3 groups which describe the symmetries of the regular and quasi regular prisms whose Coxeter diagrams are given in **Fig. 2**. Of course, the prismatic symmetries are the subgroups of the groups defined in (24). A general vertex of a quasi regular prism is given by

$$\Lambda = a_1\omega_1 + a_2\omega_2 + a_3\omega_3 = \frac{1}{\sqrt{2}}[a_1 + e_1 \frac{1}{\sin\frac{(n-1)\pi}{n}}(-a_1\cos\frac{(n-1)\pi}{n} + a_2) + a_3 e_2]. \quad (25)$$

For arbitrary values of $a_1, a_2$ and $a_3$ the vertices of the quasi regular prisms are obtained by the group action $D_n \times C_2$ on the vector $\Lambda$ in (25). However when they are all equal then the vertices of a regular prism are obtained by adding $\pm\frac{a}{\sqrt{2}}e_2$ to the vertices given in (14). Below we discuss some of the prismatic groups in turn.

### A. $n=2$ with the $D_2 \times C_2$ symmetry

Here we have the simple roots $\alpha_1/\sqrt{2} = 1$, $\alpha_2/\sqrt{2} = e_1$, $\alpha_3/\sqrt{2} = e_2$. The Coxeter group can be represented as the direct product of the cyclic groups of orders 2

$$C_2 \times C_2 \times C_2 = \{[q,\bar{q}] \oplus [q,\bar{q}]^*\}, \quad q \in Q = \{\pm 1, \pm e_1, \pm e_2, \pm e_3\}. \quad (26)$$

Here $Q$ is the quaternion group of order 8. Let us consider the orbit of the vector $\Lambda = a_1\omega_1 + a_2\omega_2 + a_3\omega_3$. The orbit will be a rectangular prism with edge lengths $a_1 \neq a_2 \neq a_3$ multiplied by $\sqrt{2}$. For $a_1 = a_2 = a_3$ we obtain a cube and for $a_1 = a_2 \neq a_3$ a square prism. Dual solids of the prisms discussed above are in general quasi dipyramids with the base being a rhombus for the case $a_1 \neq a_2 \neq a_3$ and the faces are scalene triangles. The dual of the square prism with $a_1 = a_2 \neq a_3$ is the dipyramid with the isosceles triangular faces. It is obvious that the dual of the cube is the octahedron which has a larger octahedral symmetry.

### B. $n=3$ with the $D_3 \times C_2$ symmetry

The Coxeter group here is the group $W(A_2 \oplus A_1) \approx D_3 \times C_2$. The Lie group associated with this diagram represents the group $SU(3) \times SU(2)$. It is the Standard Model of the high energy physics with the inclusion of the gauge group $U(1)$. The first orbit we wish to discuss is the

$$W(A_2 \oplus A_1)(\omega_1 + \omega_3) = \{\omega_1 \pm \omega_3, (\omega_2 - \omega_1) \pm \omega_3, -\omega_2 \pm \omega_3\}. \quad (27)$$

This is a triangular prism as shown in **Fig. 14(a).** It is vertex transitive.

Dual of the triangular prism in (27) is a dipyramid with the triangular base where the vertices are given by $\{\omega_2, \omega_1 - \omega_2, -\omega_1, \pm\frac{2}{3}\omega_3\}$. Six faces are isosceles triangles with the equal edge lengths $\sqrt{\frac{8}{9}}$ and the other is $\sqrt{2}$. It is depicted in **Fig. 14(b).**

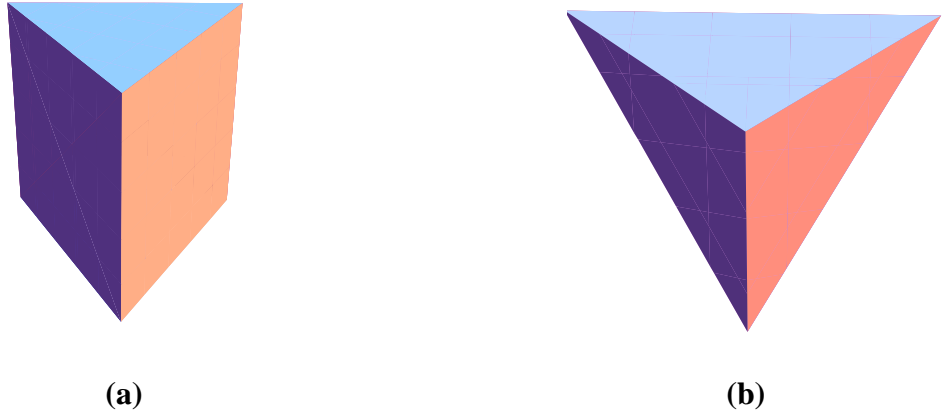

**(a)**            **(b)**

**FIG. 14. The triangular prism (a) and its dual dipyramid (b)**

The orbit $W(A_2 \oplus A_1)(\omega_2 + \omega_3)$ leads to another triangular prism which is the mirror image of the one in (27) with respect to the $x_1 = 0$ plane. The orbit $W(A_2 \oplus A_1)(\omega_1 + \omega_2 + \omega_3)$ is the hexagonal prism as shown in **Fig.15 (a).** Similarly the dual of the hexagonal prism is a dipyramid with a hexagonal base where the vertices are given by

$$\{\pm\omega_1, \pm\omega_2, \pm(\omega_2 - \omega_1), \pm 2\omega_3\}. \tag{28}$$

The faces of the dipyramid are isosceles triangles with the equal edge lengths $\sqrt{\frac{8}{3}}$ and the other is $\sqrt{\frac{2}{3}}$. It is shown in **Fig.15 (b-c).**

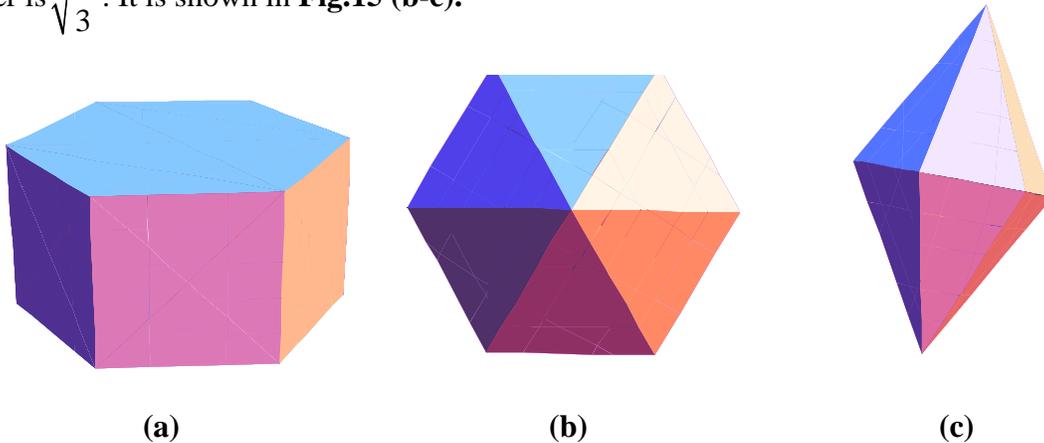

**(a)**        **(b)**        **(c)**

**FIG. 15. The hexagonal prism (a) its dual dipyramid (b) (top view) and (c) (side view)**

With a general vector $\Lambda = a_1\omega_1 + a_2\omega_2 + a_3\omega_3$ the orbit $W(A_2 \oplus A_1)(\Lambda)$ represents a quasi regular hexagonal prism consisting of two parallel faces of isogonal hexagons with two edge lengths $a_1$ and $a_2$ and three rectangular faces with edges $(a_1, a_3)$ and the three rectangular faces with edges $(a_2, a_3)$. Here we dropped a scale factor $\sqrt{2}$ in the edge lengths. The dual of this quasi regular prism is a dipyramid with isotoxal hexagonal base and scalene triangular faces. The prism and its dual with the values $a_1 \neq a_2 \neq a_3 \neq 0$ are given in **Fig.16 (a-c).**

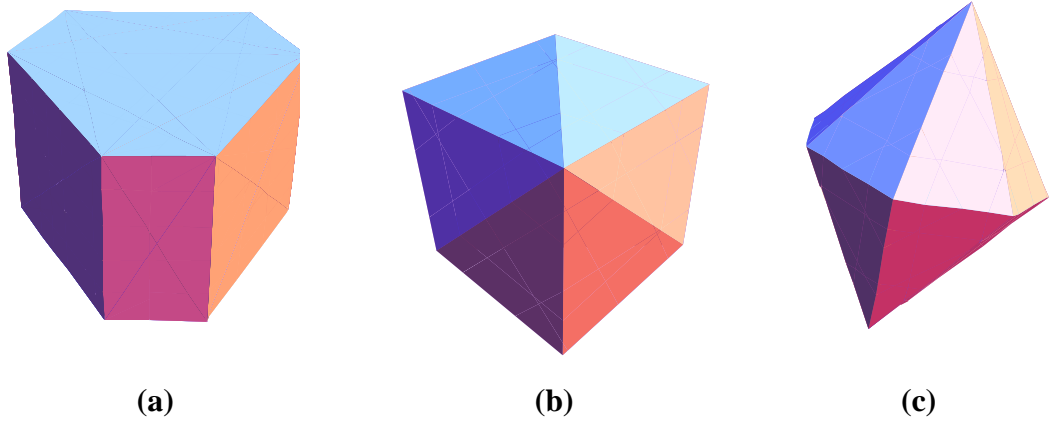

**(a)**         **(b)**         **(c)**

**FIG. 16. The isogonal hexagonal prism (a) and its dual the dipyramid with a basis of isotoxal hexagon (b) (top view) and (c) (side view)**

## C. $n=4$ with the $D_4 \times C_2$ symmetry

The Coxeter group is the $W(B_2 \oplus A_1) \approx D_4 \times C_2$ of order 16. One of the fundamental orbits is given by the set of vectors

$$W(B_2 \oplus A_1)(\omega_1 + \omega_3) = \{\pm\omega_1 \pm \omega_3, \pm(-\omega_1 + \sqrt{2}\omega_2) \pm \omega_3\} \quad (29)$$

which represents a cube. Its dual obviously is the octahedron with the vertices

$$\{\pm\omega_2, \pm(-\omega_2 + \sqrt{2}\omega_1), \pm\sqrt{2}\omega_3\}. \quad (30)$$

Similarly one another fundamental orbit is given by the set of vectors

$$W(B_2 \oplus A_1)(\omega_2 + \omega_3) = \{\pm\omega_2 \pm \omega_3, \pm(-\omega_2 + \sqrt{2}\omega_1) \pm \omega_3\} \quad (31)$$

which is a rotated cube by $45^0$ around the $\omega_3$-axis with respect to the first cube.

The orbit $W(B_2 \oplus A_1)(a_1\omega_1 + a_2\omega_2 + a_3\omega_3)$ is a prism with two parallel isogonal octagons with scaled edge lengths $(a_1, a_2)$, four rectangular faces with edges $(a_1, a_3)$ and four rectangular faces with edges $(a_2, a_3)$. Note that when $a_1 = a_2 = a_3$ the octagonal prism is uniform meaning that all the edges are equal. The dual of the uniform octagonal prism is the dipyramid with isosceles faces and regular octagonal base.

## D. $n=5$ with the $D_5 \times C_2$ symmetry

The Coxeter group is the $W(H_2 \oplus A_1) \approx D_5 \times C_2$ of order 20. Of two orbits, $W(H_2 \oplus A_1)(\omega_1 + \omega_3)$ and $W(H_2 \oplus A_1)(\omega_2 + \omega_3)$ each represents a pentagonal prism, one is rotated with respect to the other by $36^0$ around the $\omega_3$-axis. The dual of a pentagonal antiprism is the dipyramid with a pentagonal base and 10 isosceles triangles. Two equal edges are of length $\dfrac{\tau}{\sqrt{2(2+\sigma)}}$ and the third one is $\sqrt{2}$. Similar arguments can be used to discuss the quasi regular prism with isogonal parallel decagons and rectangular faces.

## V. CONCLUSION

We have displayed a method to construct the quasi regular polygons and their duals as well as the quasi regular prisms with their duals using the 2D and a subset of 3D Coxeter diagrams respectively. The isogonal polygons with *2n* sides and their duals isotoxal polygons possess the dihedral symmetry. We have shown that the aperiodic tiling of the plane with two tiles, one is the isogonal or isotoxal polygon and the second is another tile depending on the symmetry of the quasiregular polygon can be constructed. It is tempting to suggest that the tiling of the plane by the isogonal hexagon and regular hexagon may represent a state of graphene where double bond and single bonds can be represented by two edges of the isogonal hexagons. We have constructed a number of isogonal polygons with their duals possessing various dihedral symmetries .The corresponding aperiodic tilings of the plane with the tiles chosen as isogonal and isotoxal polygons are studied.

The Coxeter symmetries $D_n \times C_2$ are used to construct the regular as well as quasi regular prisms. Their duals, the quasiregular dipyramids are also constructed. We have also pointed out the correspondence between the Coxeter symmetries and the finite subgroups of the quaternions.

**Acknowledgement**

We thank Dr R Shah for helping for some of the drawings.